\documentclass[fleqn,usenatbib]{mnras}

\usepackage[T1]{fontenc}
\usepackage{listings}
\DeclareRobustCommand{\VAN}[3]{#2}
\let\VANthebibliography\thebibliography
\def\thebibliography{\DeclareRobustCommand{\VAN}[3]{##3}\VANthebibliography}

\usepackage{graphicx}	
\usepackage{amsmath}	
\usepackage{amssymb}	

\usepackage{newtxmath}

\usepackage{float}
\usepackage{subfigure}
\usepackage{multirow}

\usepackage{algorithmic}
\usepackage{algorithm}

\title[WS-Snapshot algorithm]{WS-Snapshot: An effective algorithm for wide-field and large-scale imaging}

\author[Xie et al.]{
Yang-Fan Xie,$^{1,2}$
Feng Wang,$^{1,2}$\thanks{E-mail: fengwang@gzhu.edu.cn}
Hui Deng,$^{1,2}$\thanks{E-mail: denghui@gzhu.edu.cn}
Ying Mei,$^{1,2}$\thanks{E-mail: meiying@gzhu.edu.cn}
Ying-He Celeste Lü,$^{3}$
Gabriella Hodosán,$^{4}$
\newauthor
Vladislav Stolyarov,$^{3,5}$
Oleg Smirnov,$^{6,7}$
Xiao-Feng Li,$^{1,2}$\thanks{E-mail: lixf@gzhu.edu.cn}
and Tim Cornwell$^{3}$
\\
$^{1}$ Center For Astrophysics, Guangzhou University, Guangzhou 510006, P.R. China\\
$^{2}$ Great Bay Center, National Astronomical Data Center, Guangzhou, Guangdong, 510006, P.R. China\\
$^{3}$ Cavendish Astrophysics Group, University of Cambridge, Cambridge, CB3 0HE, UK\\
$^{4}$ RAL Space, STFC Rutherford Appleton Laboratory, Didcot, Oxfordshire, OX11 0QX, UK\\
$^{5}$ Special Astrophysical Observatory of RAS, Nizhny Arkhyz, 369167, Russia\\
$^{6}$ Department of Physics and Electronics, Rhodes University, PO Box 94, Makhanda 6140, South Africa\\
$^{7}$ South African Radio Astronomy Observatory, 
Observatory 7925 South Africa\\
}

\date{Accepted XXX. Received YYY; in original form ZZZ}

\pubyear{2022}

\begin{document}
\label{firstpage}
\pagerange{\pageref{firstpage}--\pageref{lastpage}}
\maketitle

\begin{abstract}
The Square Kilometre Array (SKA) is the largest radio interferometer under construction in the world. The high accuracy, wide-field and large size imaging significantly challenge the construction of the Science Data Processor (SDP) of SKA. We propose a hybrid imaging method based on improved W-Stacking and snapshots. The w range is reduced by fitting the snapshot $uv$ plane, thus effectively enhancing the performance of the improved W-Stacking algorithm. We present a detailed implementation of WS-Snapshot. With full-scale SKA1-LOW simulations, we present the imaging performance and imaging quality results for different parameter cases. The results show that the WS-Snapshot method enables more efficient distributed processing and significantly reduces the computational time overhead within an acceptable accuracy range, which would be crucial for subsequent SKA science studies.

\end{abstract}




\section{Introduction}
\label{sec:introduction}

For a wide-field radio interferometer such as the Square Kilometre Array (SKA), imaging by a fast Fourier transform would lead to significant errors in regions far from the centre of the field of view (FOV) due to non-coplanar baseline effects~\citep{2017isra.book.....T,1990SPIE.1351..706C,RN273}. 
The relation between the visibility data $V(u,v,w)$ and the sky brightness distribution $I(l,m)$ is given by 
\begin{equation}
    V(u,v,w)=\int \frac{I(l,m)}{\sqrt{1-l^2-m^2}} e^{j2\pi(ul+vm+w(\sqrt{1-l^2-m^2}-1))} dldm.
	\label{eq:v-i}
\end{equation}
where $(l,m)$ are direction cosines and $(u,v,w)$ are baseline coordinates in units of wavelength.
Obviously, the w-term cannot be ignored when the field of view is large~\citep{1971PhDT.......153B,2017isra.book.....T,cornwell2012wide}. It is necessary to consider the effect of w-term in large fields of view in order to make the results of wide-field imaging more accurate \citep{RN274}. 

Existing wide-field imaging algorithms are dedicated to correcting the phase errors introduced by the w-term. \cite{RN274} first introduced the 3D-FFT algorithm to direct imaging, but this algorithm requires significant computational resources. The polyhedron imaging method, or image-plane faceting, was introduced to decrease computational costs. It is a more cost-effective alternative to the 3D-FFT, which has been used for wide-field imaging~\citep{RN274}. A variation of this approach, using coplanar rather than polyhedral facets, was proposed by \citet{aipsfacets}.

As discussed in ~\cite{cornwell2012wide},~\cite{ord2010interferometric},~\cite{RN273},~\cite{1971PhDT.......153B}, and ~\cite{1984iimp.conf..177B},
the sampled points in $V(u,v,w)$ space can be considered to be in the same plane for snapshot observations. 

The W-Projection algorithm projects $V(u,v,w)$ to the plane with $w=0$, followed by a 2D-FFT to do the visibility function-to-image conversion \citep{cornwell2003w,cornwell2008noncoplanar}. 
\cite{cornwell2012wide} further presented a novel algorithm to correct $w$ in combination with W-Projection and snapshots, which is called the W-Snapshot algorithm. 

In addition to W-Projection, W-Stacking has been proposed to correct for the w-term, the main idea of which is to discretize $w$ in $V(u,v,w)$ space and finally to weight the results of each w-plane cumulatively \citep{Humphreys2011,RN276}. 

Recent work has effectviely combined the methods discussed above. In particular, the DDFacet imager \citep{tasse2018faceting} combines the facet-based approach with W-projection. It employs coplanar facets, and uses a per-facet version of a modified W-kernel to image each facet using W-projection. Recent versions of the WSClean imager \citep{RN276} also implement facet-based imaging, but rather using W-Stacking. This has complex computational trade-offs: the use of physically smaller (than the image) facets allows for smaller W-kernels or fewer W-Stacks, which offsets the computational cost of imaging a large number of facets (which in itself is an embarrassingly parallel problem, allowing for very efficient parallel implementations). Both packages support on-the-fly baseline-dependent averaging (BDA), which further reduces comptational load. In the case of DDFacet, the BDA is done on a per-facet basis, thus allowing for very aggressive averaging factors. The facet approach also naturally lends itself to application of direction-dependent effects (DDEs) on a per-facet basis. Alternatively, the AW-projection technique proposed by \citet{Aproj1} combines W-projection with per-baseline gridding kernels, which also serves to correct for DDEs. A full comparison of these approaches, as well as treatment of DDEs, is outside the scope of this paper. They do, however, signpost the way to how different imaging approaches may be combined.

\cite{ye2020optimal} proposed an improved W-Stacking (IW-Stacking) algorithm that achieves very high accuracy with fewer w-planes. 
The gridding is performed in three dimensions, along with the w-axis and the usual u- and v-axes. A 'least-misfit gridding functions' kernel is presented to produce images closer to the direct Fourier transform (DFT) than the spheroidal convolution kernels traditionally used. The number of w-planes is determined by
\begin{equation}
N_{w} \geq \frac{max_{l,m} (1-\sqrt{1-l^2-m^2})(w_{max}-w_{min})}{x_0} +W
\label{eq:wplane}
\end{equation}
where $(w_{max}-w_{min})$ is the w-range. $(1-\sqrt{1-l^2-m^2})$ is determined by the FOV. $W$ is the support of the gridding kernel function. $x_{0}$ is a parameter to control the retained portion of the image. In addition, to obtain the best results, all images are recommended to make twice as large as necessary in each dimension, and the outer half cropped and discarded while using this kernel.

Based on the IW-Stacking algorithm, \cite{RN278} implemented efficient and accurate gridding and degridding operators, i.e., Ducc.wgridder\footnote{https://gitlab.mpcdf.mpg.de/mtr/ducc}. The Ducc.wgridder can solve the w-term effect with high accuracy (down to $\approx 10^{-12}$) and performs well in terms of computation time and memory consumption in practice. Ducc.wgridder is written in C++ and supports multiple threads to improve performance.
Due to its outstanding processing performance and accuracy, Ducc.wgridder quickly became a vital software package and has been widely used in radio interferometer imaging. 

However, in the case of large FOVs and large image sizes, e.g., the 10 square degrees of FOV in continuum imaging of SKA1-Low and the image size of  $2^{15}$ pixels \citep{cornwell2012wide,scaife2020big},  we noticed deficiencies while using IW-Stacking (Ducc.wgridder) for wide-field imaging. 
We tested the performance with a range of threads from 1 to 40 and different scales such as 4K $(4096\times4096)$, 8K $(8192\times8192)$, and 16K $(16384\times16384)$, respectively. 

\begin{figure}
	\includegraphics[width=\columnwidth]{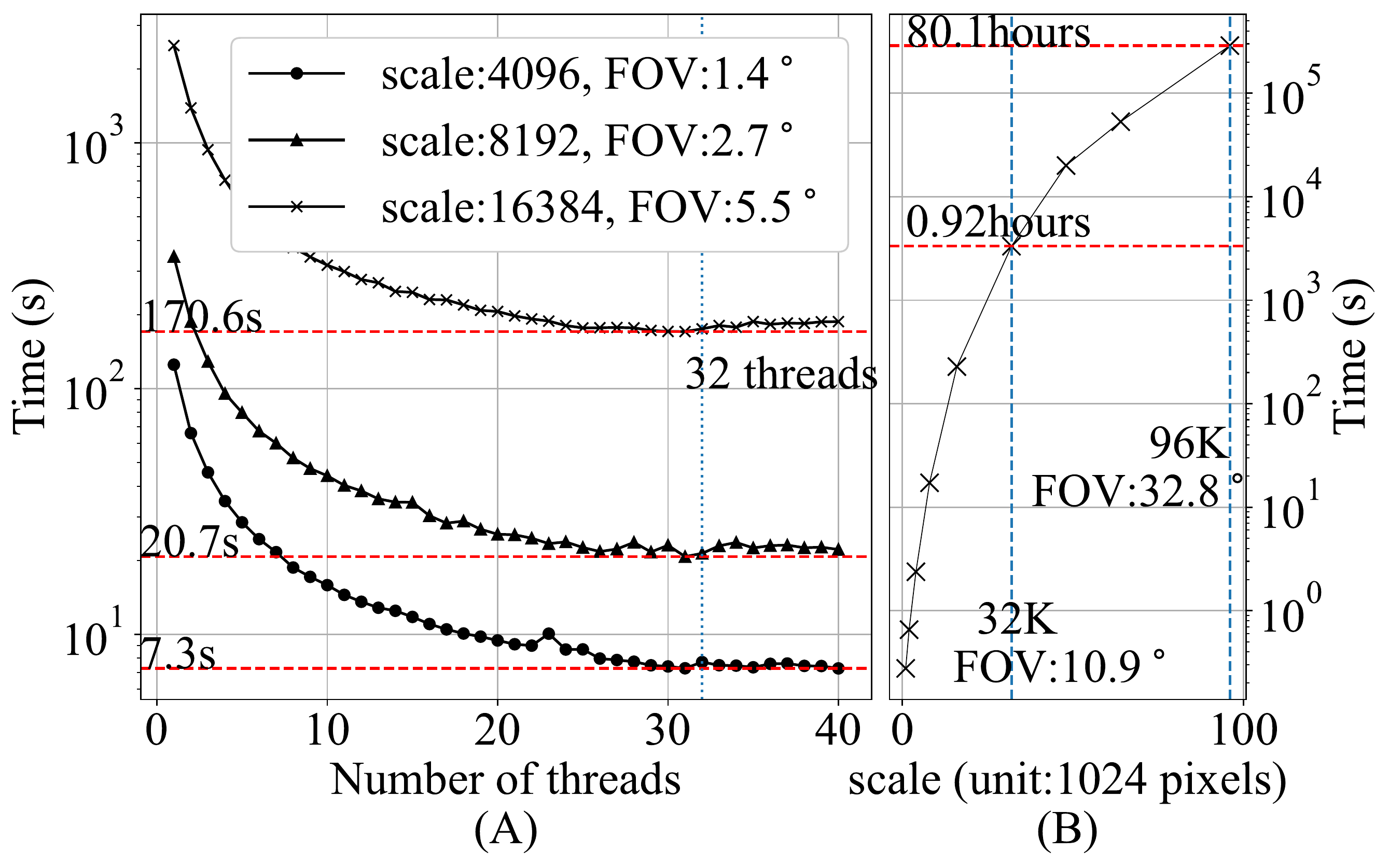}
    \caption{(A): The diagram of time consumption of IW-Stacking (Ducc.wgridder) with different number of threads. (B): The diagram of time consumption of IW-Stacking (Ducc.wgridder) in different image scales running at a 32-core server.}
    \label{fig:nglimt}
\end{figure}

The test results are shown in Figure \ref{fig:nglimt}. When imaging calculation is performed on large images, there is no longer any way to increase the final processing speed by increasing the number of threads, which is a limitation of IW-Stacking for SKA large-scale data processing.

To meet the requirements of large-scale imaging, it is worth improving processing performance as much as possible while ensuring sufficient imaging accuracy. Reducing the number of w-planes is a possible method to reduce calculation time. Snapshot imaging allows a significant reduction of w-range by coordinate transformation based on splitting visibility data into multiple snapshot observations  \citep{1971PhDT.......153B,1984iimp.conf..177B,cornwell2012wide,ord2010interferometric}. To fully take advantage of the excellent performance of "improved W-Stacking", we propose a hybrid algorithm WS-Snapshot that has the advantages of both "improved W-Stacking" and snapshot algorithms. 

We introduce the WS-Snapshot algorithm in Section \ref{sec:WS-Snapshot algorithm}, investigate its imaging accuracy and performance through practical tests in Section \ref{sec:performance evaluation}, and discuss some limitations of WS-Snapshot in Section \ref{sec:discussion}.

\section{Distributed WS-Snapshot Algorithm}
\label{sec:WS-Snapshot algorithm}
\subsection{WS-Snapshot}

\cite{cornwell2012wide}  showed that the $w$-term can be expressed as a linear plane plus deviations $\Delta w$ when the antenna array is not strictly co-planar. The $w$-term is determined by 
\begin{equation}\label{eq_uvw}
w=a u+b v +\Delta w 
\end{equation}
where $ a = \tan Z \sin \chi$, $b = - \tan Z \cos \chi$,
$Z$ is zenith distance, and $\chi$ is parallactic angle. 

WS-Snapshot extends W-Snapshot by using IW-Stacking to correct $\Delta w$ instead of W-Projection. 
We first obtain the current best plane in u,v,w space by least-squares fit. We then use IW-Stacking to correct $\Delta w$ based on the fitted plane and perform snapshot imaging. We finally re-project the results of each snapshot from a distorted coordinate system to a normal coordinate system. 
Two parameters, $a$ and $b$ (eq \ref{eq_uvw}),  can be stored as the projection parameters of the image obtained by the snapshot observation, which in the current Flexible Image Transport System (FITS) are PV2\_1 and PV2\_2 respectively~\citep{ord2010interferometric}.

\begin{figure}
	\includegraphics[width=\columnwidth]{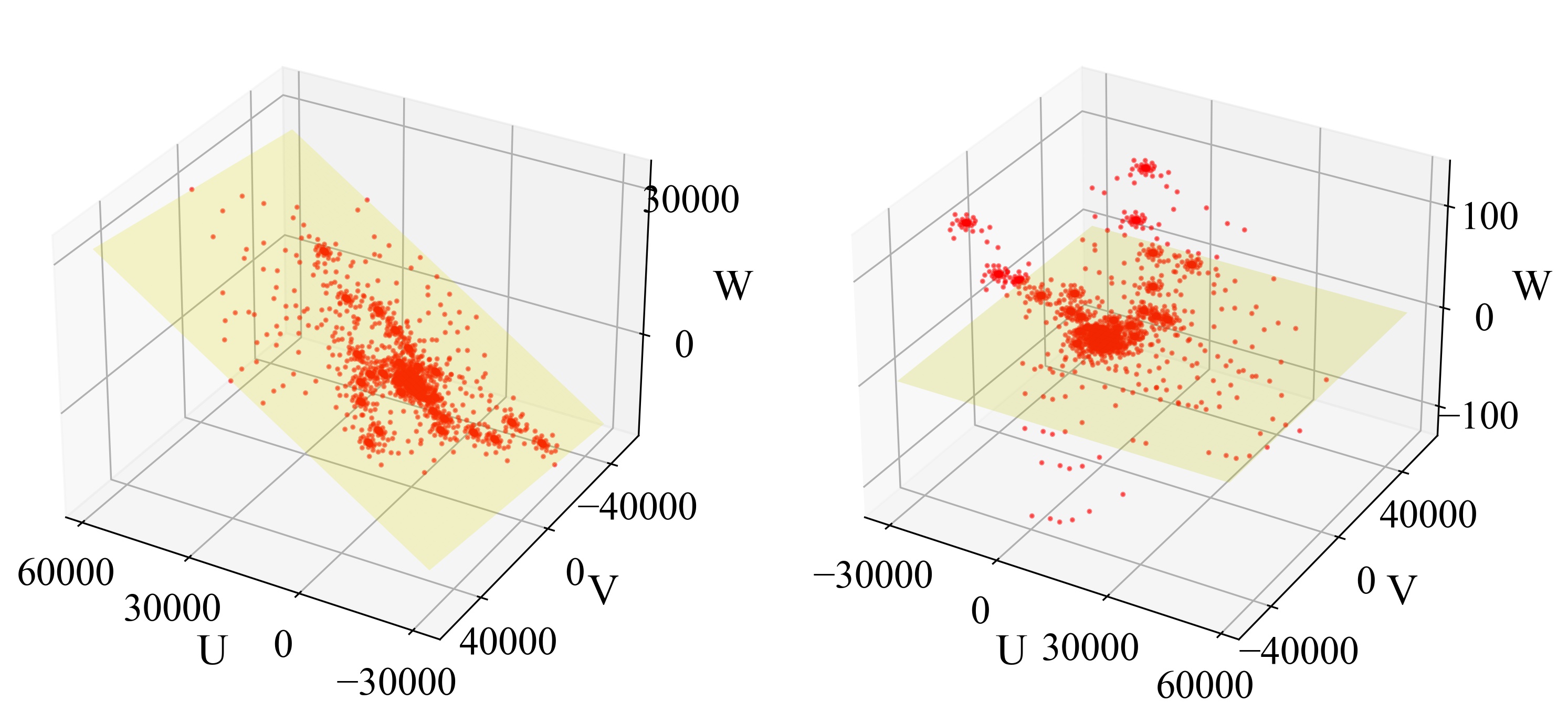}
    \caption{The diagram of uvw distribution in a snapshot. The left and right diagrams show the w-plane before (w-range = 45,511 metres) and after (w-range = 239 metres) plane fit, respectively.}
    \label{fig:fitwplane}
\end{figure}

\begin{table*}
\centering
\renewcommand{\arraystretch}{1.0}

\caption{Computational Complexity. $N_{x}$ and $N_{y}$ are the pixel size of the image in height and width respectively, FOV is the field of view, $W_{\rm range}$ and $W_{\rm range\_fit}$ are the values of w-range before and after the w-plane fitting, respectively. $N_{\rm vis}$ is the amount of visibility data. $N_{\rm fac}$ is the number of facets. $B$ is the BDA compression factor. The computational cost for FFTs and gridding is shown separately.}
\label{tab:complex}
\begin{tabular}{lll}
\hline
Method                           &FFT cost                                     &Gridding cost                  \\\hline
W-Snapshot(snapshot imaging)    &$\mathcal O(N_{x}N_{y}\log({N_{x}N_{y}}))$                                   &$\mathcal O(N_{\rm vis})$ \\
W-Stacking                       &$\mathcal O(W_{\rm range} \times FOV \times N_{x}N_{y}\log({N_{x}N_{y}}))$          &$\mathcal O(N_{\rm vis})$   \\
WS-Snapshot(snapshot imaging)   &$\mathcal O(W_{\rm range\_fit} \times FOV \times N_{x}N_{y}\log({N_{x}N_{y}}) + N_{x}N_{y})$  &$\mathcal O(N_{\rm vis})$ \\
DDFacet & $\mathcal O(N_{x}N_{y}\log({N_{x}N_{y}/N_{\rm fac}) + N_{x}N_{y}}))$  &$\mathcal O(N_{\rm vis}N_{\rm fac}/B)$ \\
\hline
\end{tabular}
\end{table*}

The computational complexities of W-Snapshot, W-Stacking, and WS-Snapshot are shown in Table \ref{tab:complex}. 
Although WS-Snapshot has $N_{x}N_{y}$ more reprojection operations than W-Stacking for a single snapshot imaging, the w-range is significantly reduced after w-plane fitting. Take the example given in Fig. \ref{fig:fitwplane}. The w-range is reduced from 45,511 m to 239 m after plane fitting, making the computational overhead required for the final WS-Snapshot to complete snapshot imaging comparable to W-Stacking or even smaller. As the FOV and image scale increase, the time consumed by both WS-Snapshot and W-Stacking increases accordingly. However, the time increase of WS-snapshot is much lower than that of W-Stacking. For reference, we also include the notional computational complexity of the DDFacet approach. Note that the number of facets $N_{\rm fac}$ can be taken to be roughly equivalent to the FOV squared. These figures should be taken with a serious caveat, as the scaling constants can be very different. In particular, the aggressive BDA strategy employed by DDFacet (see above) can easily yield compression factors of $B \gg 10$ for small facets (large $N_{\rm fac}$), but this is a complicated scaling relationship. It is also essential to clarify that WS-Snapshot is a time-slice-oriented data processing method, while DDFacet parallelizes over facets. 

\subsection{Algorithm Implementation}

We implement a WS-Snapshot function based on the Radio Astronomy Simulation, Calibration, and Imaging Library (RASCIL\footnote{https://gitlab.com/ska-telescope/external/rascil}). 
RASCIL is a crucial software package for SKA simulation and parallel data processing. RASCIL integrates with Ducc.wgridder to improve the performance of RASCIL imaging calculations.
RASCIL uses Dask \citep{rocklin2015dask} as an execution framework to build complex pipelines flexibly, which provides great convenience for WS-Snapshot implementation.

The key of the WS-Snapshot algorithm is to process multiple snapshot data separately and then combine them. Therefore, compared with other algorithms, WS-snapshot can achieve a more efficient distribution calculation by grouping the visibility data at different observation times.

To more flexibly control the parallelism of the time distributed WS-Snapshot algorithm at a limited number of nodes, two parameters, ``number of times per task (NPT)'' and ``number of times per slice (NPS)'', are considered in the implementation.
Here, the slice refers to a slice of visibility data on the time axis. A sample time can be considered a snapshot observation. A slice would include one or more snapshot observations. 
NPT means the number of snapshot observations to be processed in each computing node. The NPS defines the number of observational times in each slice. The larger NPS, the larger the w-range of a time slice after transforming the $uvw$ coordinates.

\begin{figure*}
	\includegraphics[width=0.9\textwidth]{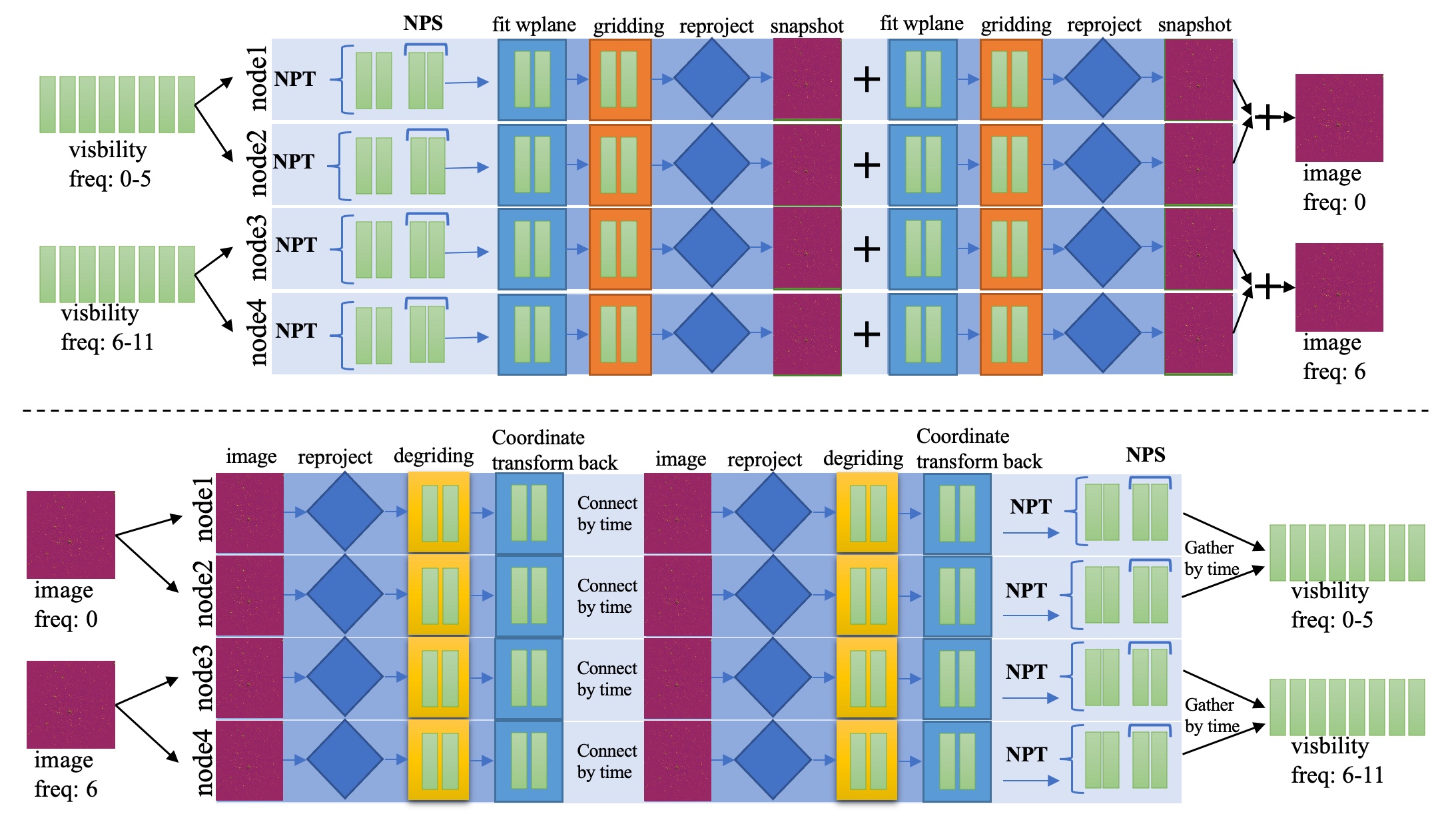}
    \caption{ The diagram of gridding/degridding processes using WS-Snapshot and MFS}
    \label{fig:ws_griding_degriding}
\end{figure*}

While processing time-distributed visibility data, WS-Snapshot is fully capable of working with multiple-frequency synthesis (MFS)~\citep{sault1994multi,conway1990multi} and multi-scale multi-frequency deconvolution algorithm (MSMFS) \citep{rau2011multi}, thus maximizing performance.
Figure \ref{fig:ws_griding_degriding} shows a diagram of WS-Snapshot gridding and degridding with MFS, respectively. The visibility data with six channels are distributed into two computing nodes when NPT=4. Ducc.wgridder has supports MFS very well, which can be directly invoked in the WS-Snapshot implementation. 
We describe the pseudo-gridding code as Algorithm \ref{alg:Griding} and the pseudo-degridding code as Algorithm \ref{alg:Degriding}

We adopt the reproject\_interp function from reproject package~\footnote{https://github.com/astropy/reproject}. The reproject package implements image reprojection (resampling) methods for astronomical images  \citep{robitaille2020reproject}. The reproject package makes it easy to perform image projection in gridding and degridding with the correct PV2\_1 and PV2\_2 FITS keywords.

\begin{algorithm}[!h]
    \caption{WS-Snapshot Gridding Pseudo-code}
    \label{alg:Griding}
    \renewcommand{\algorithmicrequire}{\textbf{Input:}}
    \renewcommand{\algorithmicensure}{\textbf{Output:}}
    \begin{algorithmic}[1]
        \REQUIRE visibility data  
        \ENSURE dirty image   
        \STATE  Divide visibility data by NPT
        \FOR{each parallel group (Distributed)}
            \STATE Divide the visibility data into slices according to NPS
            \FOR{each slice in group}
            \STATE Fit the uvw coordinates of visibility using least squares
            \STATE Use Ducc.wgridder for gridding 
            \STATE Re-project the image to normal coordinates
            \ENDFOR
            \STATE Combine the image of each slice
        \ENDFOR
        \STATE Combine the image of each group and divide by the total number of slices
        \RETURN dirty image
    \end{algorithmic}
\end{algorithm}
\begin{algorithm}[!h]
    \caption{WS-Snapshot Degridding Pseudo-code)}
    \label{alg:Degriding}
    \renewcommand{\algorithmicrequire}{\textbf{Input:}}
    \renewcommand{\algorithmicensure}{\textbf{Output:}}
    \begin{algorithmic}[1]
        \REQUIRE Sky model image  
        \ENSURE visibility data    
        \STATE  Copy model image by NPT for groups
        \FOR{each parallel group (Distributed)}
            \STATE Divide the sky model visibility data into slices according to NPS
            \FOR{each slice in group}
            \STATE Fit the uvw coordinates of model visibility data using least squares
            \STATE Re-project the model image to distorted coordinates
            \STATE Use ducc.wgridder for degridding with distorted coordinates
            \STATE Flip back the uvw coordinates of slice
            \ENDFOR
            \STATE Combine the visibility data of each slice by time
        \ENDFOR
        \STATE Combine the visibility of each group by time
        \RETURN visibility
    \end{algorithmic}
\end{algorithm}

\section{Algorithm Assessment}
\label{sec:performance evaluation}
To further evaluate the usability of WS-Snapshot, we assessed its performance with simulated data.

\subsection{Data Preparation}
We select SKA1-Low as the telescope for data simulation. All performance evaluations are based on the Measurement Set (MS) format generated by the simulated observations carried out by OSKAR\footnote{https://github.com/OxfordSKA/OSKAR} (A GPU-accelerated simulator for the Square Kilometre Array). The main simulation parameters are listed in Table \ref{tab:oskarp}. 
The final MS file for evaluation is around 17GB, 5,232,640 rows.

\begin{table}
\centering
\renewcommand{\arraystretch}{1.0}
\caption{Simulation parameters used for OSKAR}
\label{tab:oskarp}
\begin{tabular}{ll}
\hline
Parameters             & Values                     \\\hline
skymodel-1               & GLEAM \\
skymodel-2               & random faint source (FOV$< 5^\circ$) \\
fainter source         & Minimum flux $10^{-4}$ Jy      \\
telescope              & SKA1-LOW full array       \\
phase\_centre          & RA: $0.0^{\circ}$, DEC: $-27.0^{\circ}$ \\
start\_frequency\_hz   & 140 MHz                     \\
num\_channels          & 30                        \\
frequency\_inc\_hz     & 200 kHz                  \\
channel\_bandwidth\_hz & 10  kHz                      \\
observation length     & 14,400 s                    \\
num\_time\_steps       & 40                        \\
time\_average\_sec     & 0.9 s                      \\ \hline
\end{tabular}
\end{table}

\subsection{Evaluation Environment}

The evaluations were conducted on a small high-performance computing cluster with 13 nodes. All nodes have an Intel(R) Xeon(R) Gold 6226R CPU (32-core/64-thread) with 1024GB memory and are connected by a 100Gbps Infiniband network (EDR). 
All computer nodes were installed with Centos 7.9 Linux operating system with the newest updates, and Slurm 20.04 software was installed for task scheduling.

To investigate WS-Snapshot's imaging performance and quality, we utilized a RASCIL-based continuum imaging pipeline (CIP) for the SKA telescope. IW-Stacking (Ducc.gridder) has been integrated into the CIP to perform the gridding/degridding, and Fourier transform. 

All evaluation scripts are written in python3, based on RASCIL version 0.5.0, Ducc.wgridder version 0.21.0, and Dask version 2022.1.0.

\subsection{Computational Performance}

\subsubsection{WS-Snapshot vs. IW-Stacking}

\begin{table}
\centering
\renewcommand{\arraystretch}{1.0}
\caption{Scaling Parameters}
\label{tab:scalingpar}
\begin{tabular}{ll}
\hline
Parameters             & Values                     \\\hline
Scale                  & 16K($5.5^\circ$), 24K($8.2^\circ$), 32K($10.9^\circ$) \\
NPS                    & 1, 4, 8, 20 \\
Number of threads                & 8, 16, 24, 32, 40 \\
MFS channel            & 6 (140MHz-141MHZ) \\ \hline
\end{tabular}
\begin{tabular}{l}
NPS=12,16 are not divisible by 40 times and are not used in test
\end{tabular}
\end{table}

We first investigate the performance of the WS-Snapshot algorithm and analyze it in comparison with the IW-Stacking algorithm. The parameters evaluated in detail are shown in Table~\ref{tab:scalingpar}. To accurately evaluate the performance in the case of slicing at different observation times, we tested the cases of NPS equal to 1, 4, 8, and 20, respectively. For each of the  NPS values, we evaluated the imaging performance at different scales such as 16K ($16,384\times16,384$), 24K ($24,576\times24,576$) and 32K ($32,768\times32,768$). 

\begin{figure*}
	\includegraphics[width=\textwidth]{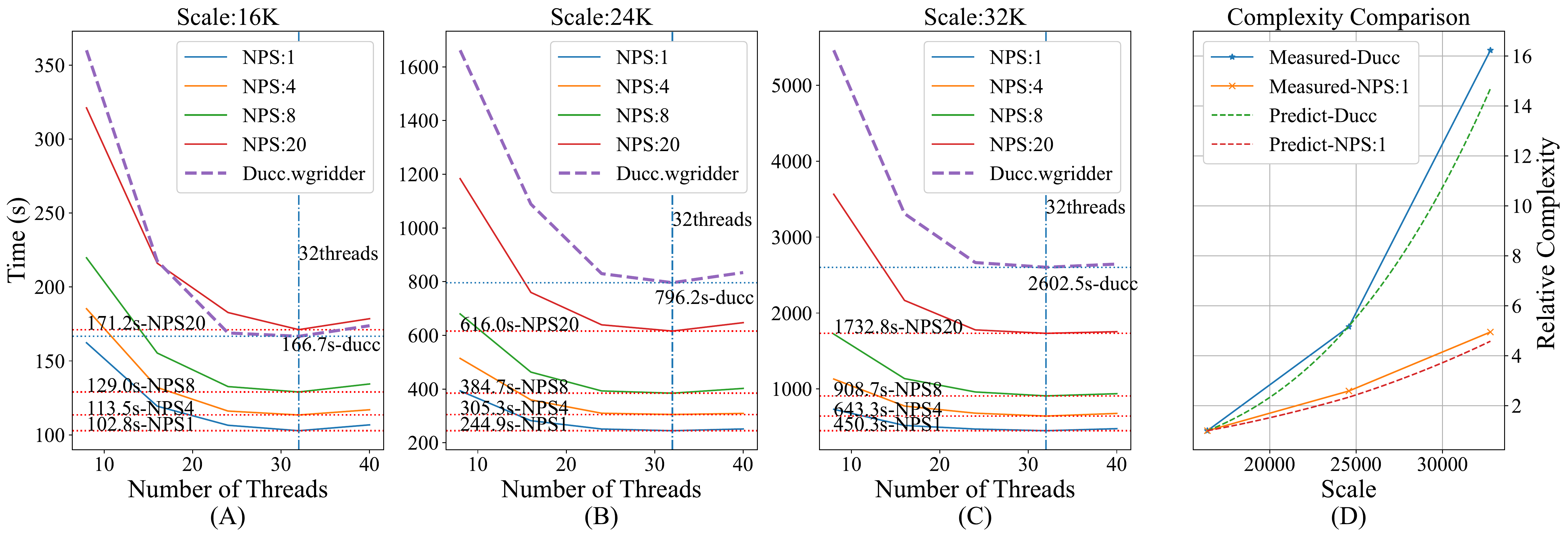}
    \caption{A, B and C are the diagram of time consumption of WS-Snapshot and IW-Stacking (Ducc.wgridder) for different number of threads; D shows the relative complexity for Ducc and WS-Snapshot (NPS=1) predictions and measurements as function of scales} 
    \label{fig:per_ws}
\end{figure*}

The final evaluation results are given in Figure~\ref{fig:per_ws}. The time consumption of IW-Stacking is not related to the number of time slices but only to the image size. The time consumption of IW-Stacking gradually decreases as the number of threads grows, but the change is approximately stable after using 24 threads.

The time consumption variation of the WS-Snapshot algorithm is approximately the same as IW-Stacking. However, regardless of the value of NPS, most of the WS-Snapshot time consumption is better than IW-Stacking, and the trend of WS-Snapshot performance with the number of threads is also basically the same as IW-Stacking. 
WS-Snapshot significantly outperforms IW-Stacking for small NPS values and large scales. Regardless of the image size, a smaller NPS will significantly reduce the computational time,
e.g., the minimum time consumption of WS-Snapshot is 450.3s at 32K with NPS=1, and the minimum time consumption IW-Stacking is 2602.5s. This means that WS-Snapshot is easier to achieve fine-grained parallelism with a sufficient number of compute nodes.

We tested the computation time of Ducc.wgridder and WS-Snapshot with different scales (i.e., 16K, 24K, 32K), and then divided these computation times by the computation time of 16K to obtain the computation time ratio. Similarly, we used the computational complexity equation in the Table \ref{tab:complex} to predict 100 computational complexities within the scale from 16K to 32K. We also calculated the ratio of the predicted values to the predicted values for 16K complexity. Figure~\ref{fig:per_ws} (D) shows that the change in the ratio of computation time to prediction complexity is essentially the same. This also indicates that the complexity estimates are reasonable.

\subsubsection{Comprehensive Performance Assessment}

\begin{table}
\centering
\renewcommand{\arraystretch}{1.0}
\caption{CIP test Parameters}
\label{tab:cippar}
\begin{tabular}{ll}
\hline
Parameters             & Values                     \\\hline
scale                  & 16K($5.5^\circ$), 24K($8.2^\circ$), 32K($10.9^\circ$) \\
NPS                    & 4 \\
number of taylor                & 2 \\
clean threshold        & $1.2 \times 10^{-3}$ Jy (>10 times minimum brightness)\\
major cycle            & 10 \\
minior cycle           & 10,000 \\
Input channels         & 30 \\
MFS channel            & 6 \\ 
number of theads       & 16 (for gridding/degridding) \\
number of Dask-worker  & 52 (13-nodes, 4 worker per node)\\\hline
\end{tabular}
\begin{tabular}{l}
The number of channels in the dataset is 30, generate one image for every 6 \\
channels, output of 5 channels, represented by 2 taylor images
\end{tabular}
\end{table}

To evaluate the WS-Snapshot algorithm more objectively, we performed a comprehensive assessment of WS-Snapshot and IW-Stacking using the continuous imaging pipeline of SKA1-LOW. Assessment parameters such as different imaging sizes, number of threads, and number of time samples per slice were considered in the evaluation (see Table ~\ref{tab:cippar}).
Likewise, we compared the imaging performance at different scales separately. 

During the assessment, we create 52 ``Dask-workers'' on 13 computing nodes. Each node runs  4 ``Dask-workers''. ``Dask-scheduler'’ is running on the first node of the cluster simultaneously.

\begin{figure}
	\includegraphics[width=\columnwidth]{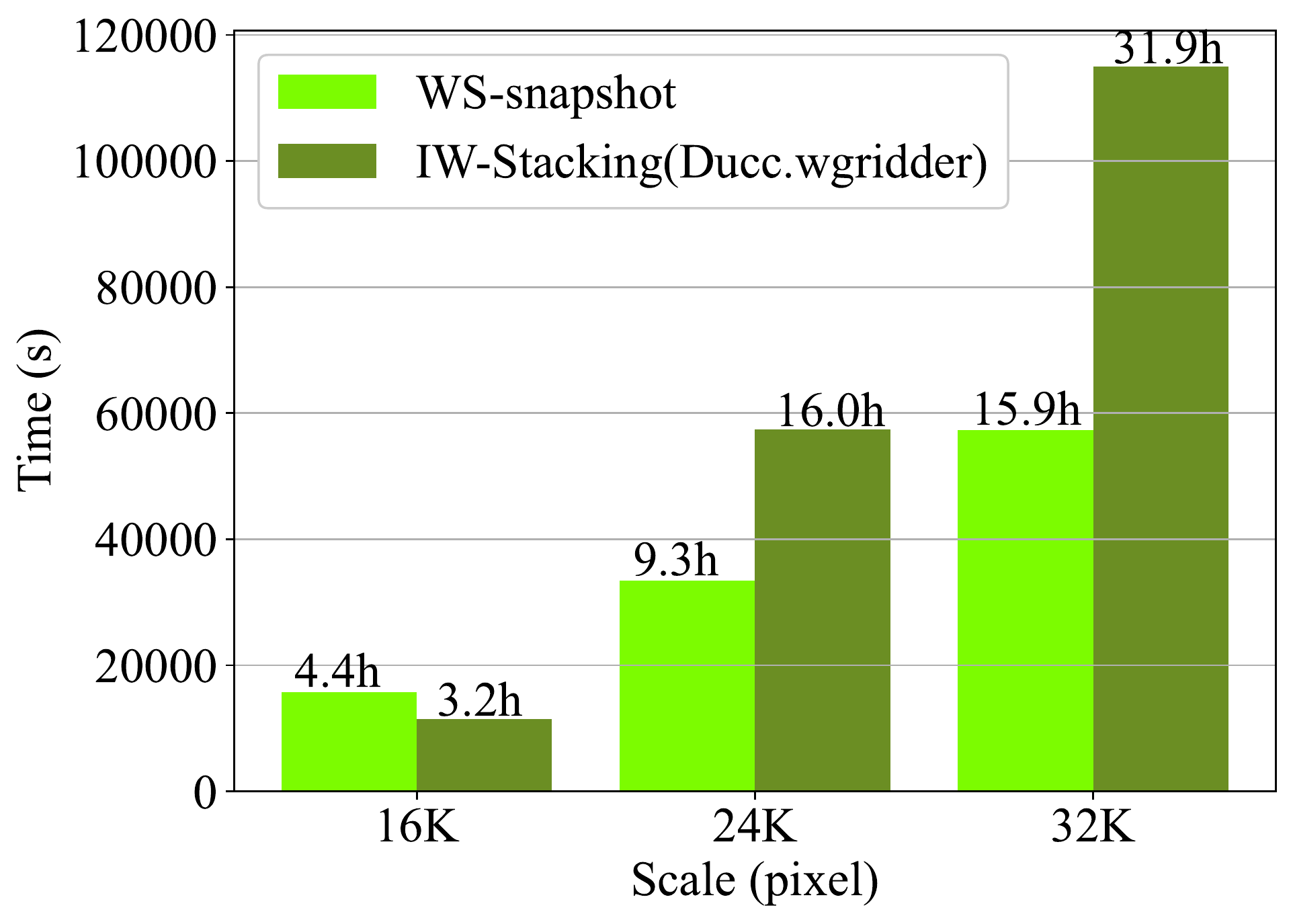}
    \caption{Histogram of CIP time consumption at different scales based on WS-Snapshot and IW-Stacking}
    \label{fig:cipper}
\end{figure}

Figure \ref{fig:cipper} shows that compared with IW-Stacking (Ducc.wgridder), WS-Snapshot has a significant improvement in the combined processing performance, especially at the scale of 32K, WS-Snapshot can save nearly half of the time.

\subsection{Imaging Quality Assessment}

We carefully analyzed the final imaging quality of the WS-Snapshot algorithm. We assessed the imaging quality in terms of both calculated dirty image and CIP results, using IW-Stacking imaging results as a comparison.

\subsubsection{Imaging Quality Assessment}

\begin{figure}
	\includegraphics[width=\columnwidth]{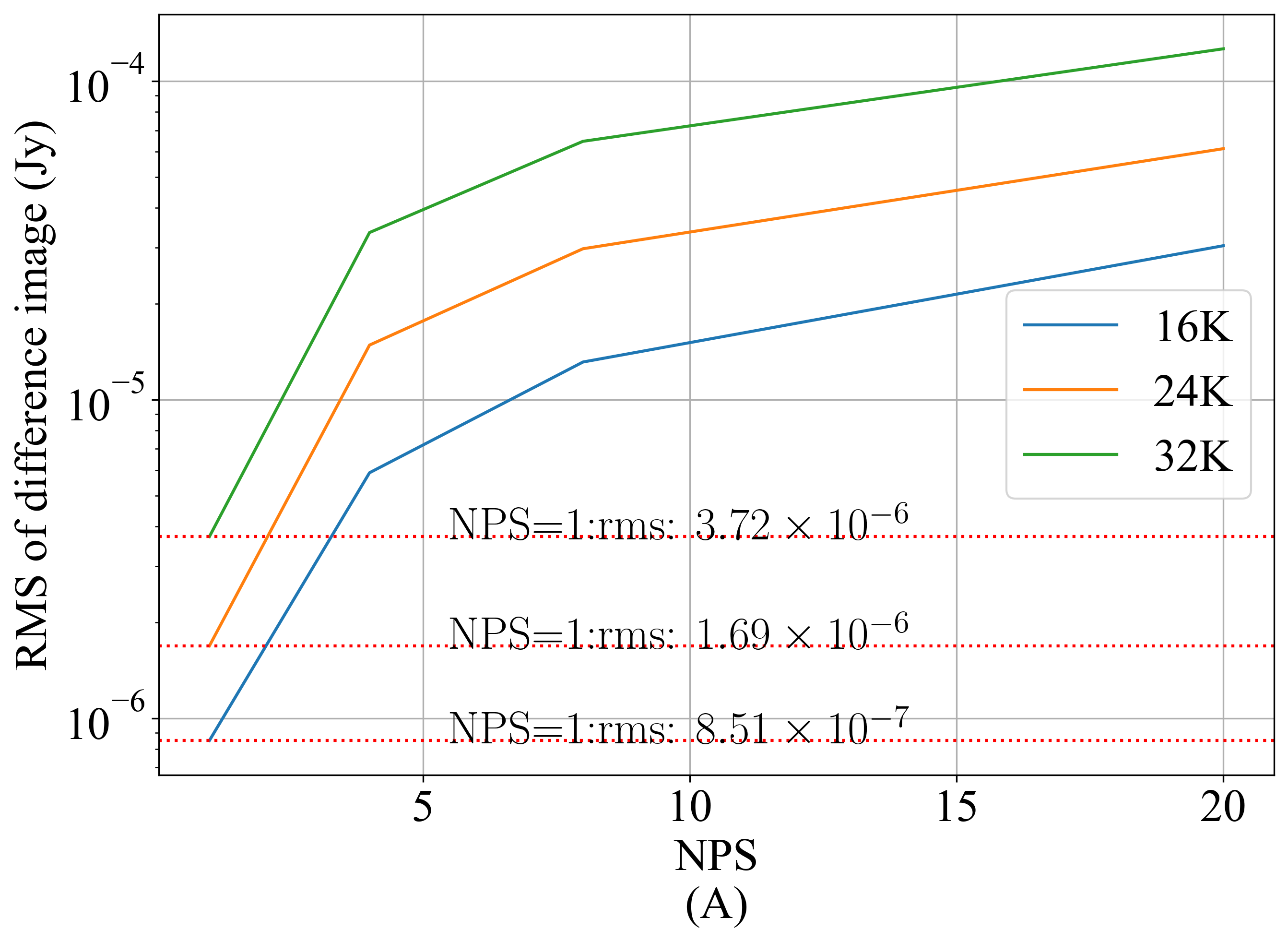}
    \caption{the RMS variations with different NPS.}
    \label{fig:per_error_wplane}
\end{figure}

We subtracted the dirty images generated by WS-Snapshot and IW-Stacking at the same scale and NPS to obtain the difference images. The root means square error (RMS) is calculated for the difference images. The RMS for different scales with different NPS are given in Figure~\ref{fig:per_error_wplane}.

As the NPS value increases, the RMS value increases from $10^{-6}$ to $10^{-4}$. As expected, larger imaging scales or larger NPS values will result in more significant errors.

\begin{figure*}
	\includegraphics[width=\textwidth]{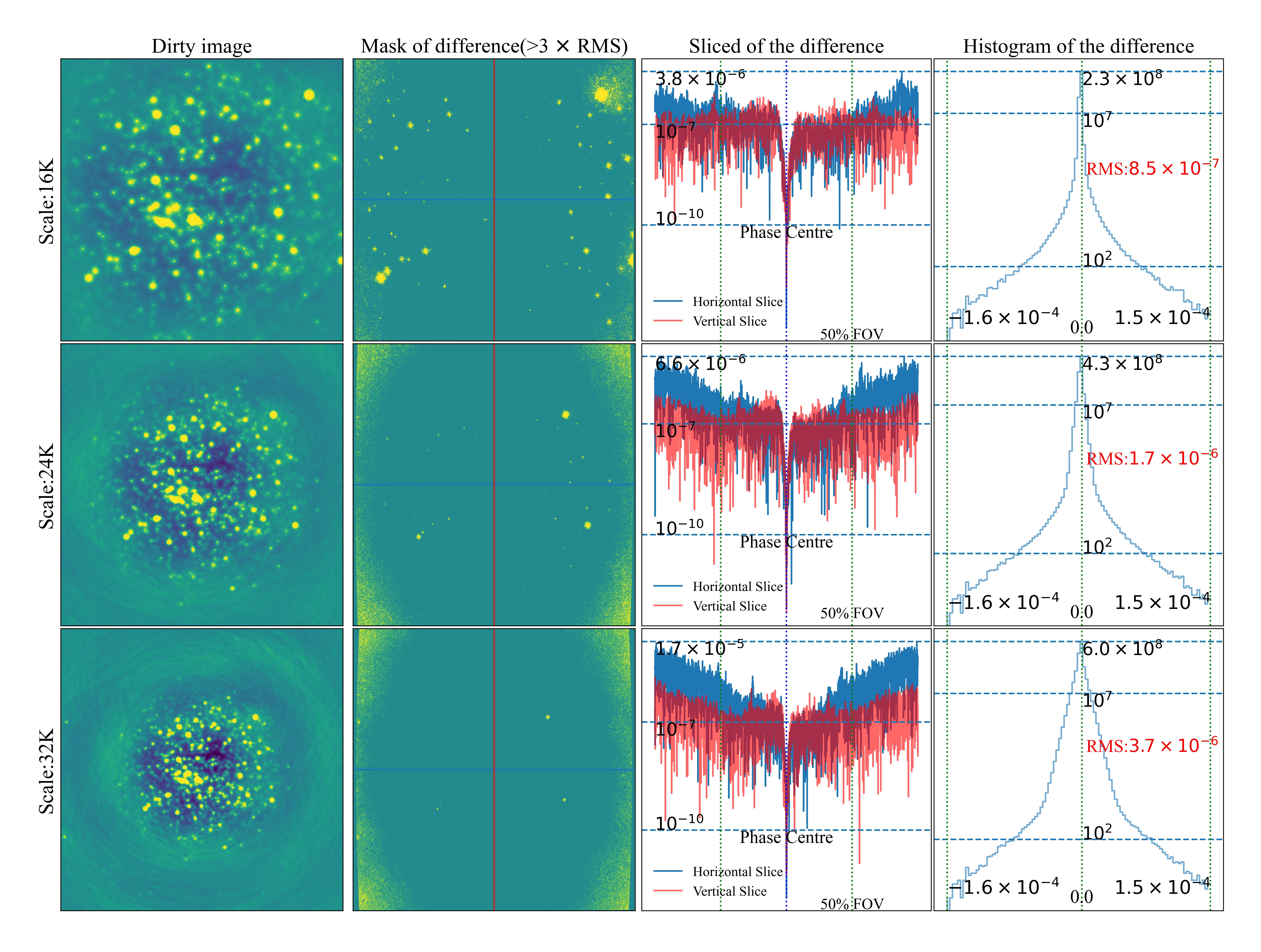}
    \caption{The dirty and the difference images and histograms of the difference are shown here for NPS of 1, three scales, with the horizontal and vertical slices of the difference image given, and the vertical coordinates of the difference slices and histograms are taken as log.}
    \label{fig:wsimage}
\end{figure*}

To further assess the imaging quality, we set the NPS to 1, performed imaging calculations on the 40 observation times of visibility data included in the simulated MS, and then superimposed the generated 40 dirty images to generate the final dirty image (see Figure~\ref{fig:wsimage}). 
We further used the generated dirty image to subtract the dirty image produced by IW-Stacking to obtain difference images.

Due to the limited re-projection accuracy,  the RMS of the edges is much higher than that of the central region in the difference image. It can also be seen from the histogram of the RMS that the central part of the image should be selected for the final image when high precision imaging is required.

\subsubsection{Quality Assessment of CIP Results}

\begin{figure*}
	\includegraphics[width=0.8\textwidth]{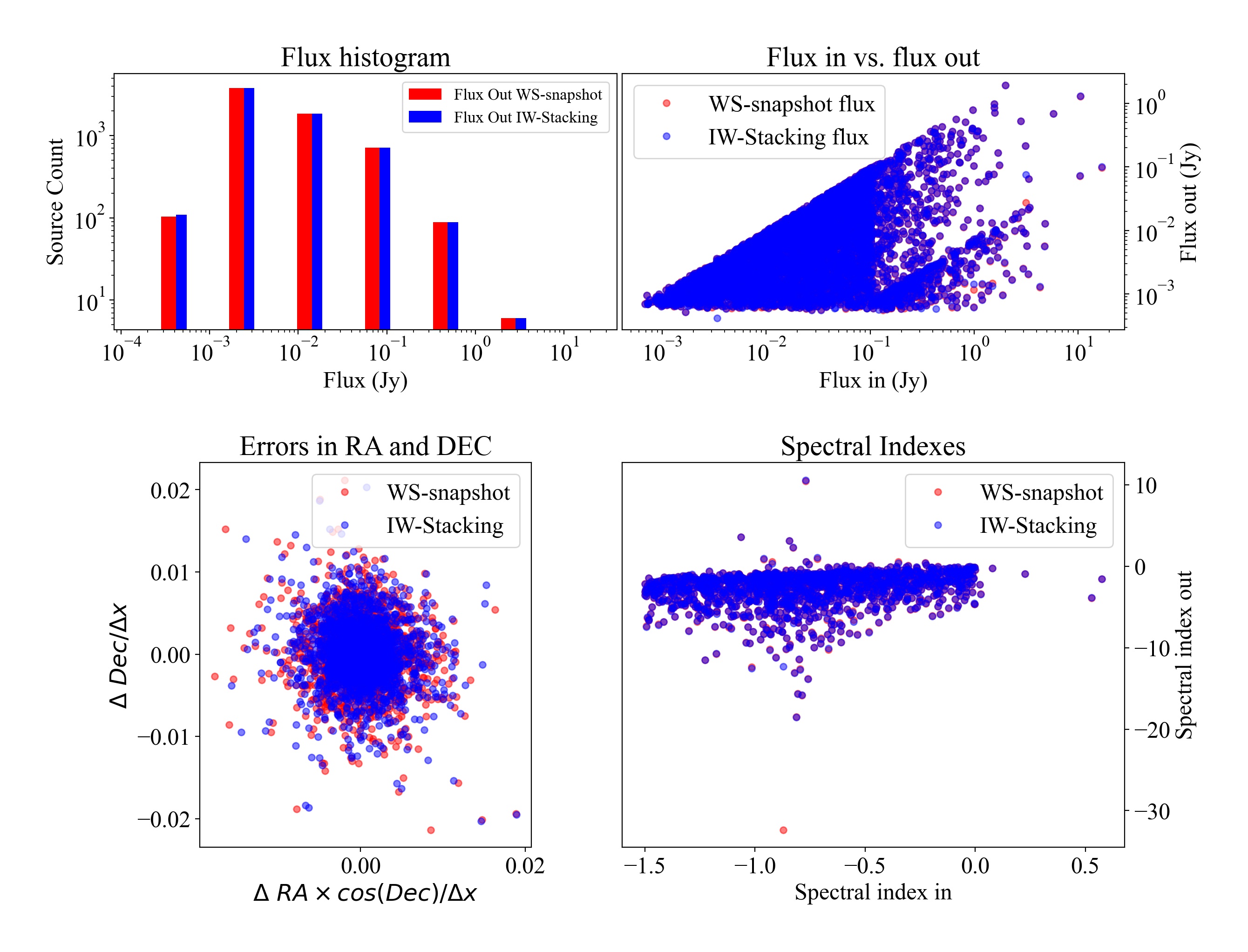}
    \caption{
    Comparison of WS-Snapshot and IW-Stacking(Ducc.wgridder) as gridder for CIP imaging quality evaluation at 32K pixels 10 degree FOV. Up left: Histogram of the fluxes of all sources, binned logarithmically, where red and blue represents the output catalogue with two methods; Up right: The comparison of fluxs of all sources for input and output source catalogues; Bottom left: The errors between identified and input source positions, with respect to image resolution in (RA, Dec); Bottom right: The comparison of spectral indexes for input and output source catalogues.}
    \label{fig:cipqa}
\end{figure*}

\begin{figure*}
	\includegraphics[width=\textwidth]{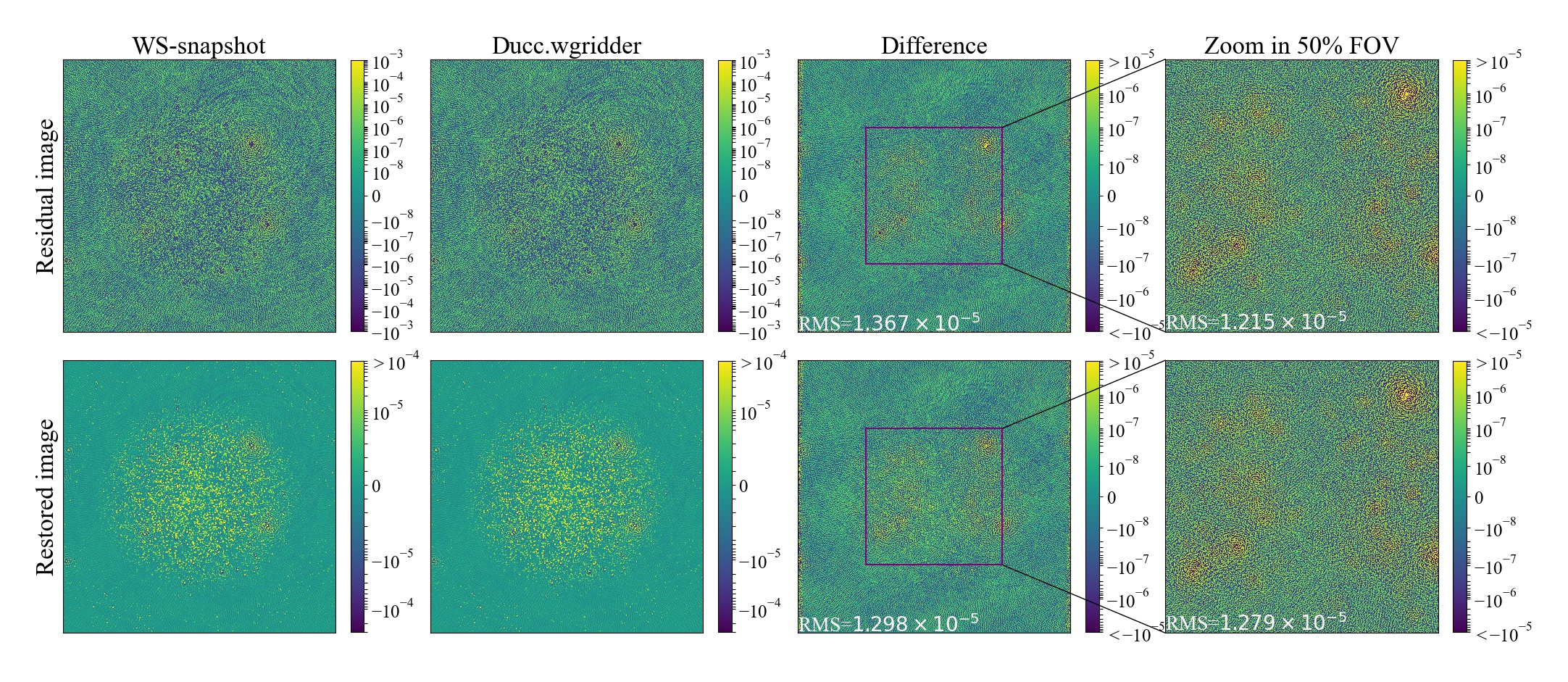}
    \caption{IW-Stacking (Ducc.wgridder) and WS-Snapshot are respectively applied to the 0th-order Taylor residuals image and the restored image of the CIP, and the corresponding difference (npixel:32K,FoV:10.9 degrees). Also shown here is a partial zoom of the snapshot dirty image of the bright source.}
    \label{fig:cipres}
\end{figure*}

To objectively compare the imaging quality, we evaluated the imaging results of both algorithms separately using the quality assessment (QA) procedure of the RASCIL QA tool \citep{ying2022qa}, and the results are shown in Figure~\ref{fig:cipqa}. 
With the default parameters of RASCIL QA, the input fluxes are absolute and the output fluxes are apparent.
The three results show that: 1. the source coordinate positions obtained by QA through the source search method are approximately the same, with differences in 2\% of error in both directions compared to the input source catalogue. 2. the spectral index of a source varies widely, with relatively small differences in brightness; 3. the number of WS-Snapshot weak source detections is statistically consistent with IW-Stacking (within $10^1$).

From the final imaging results, the 0th-order Taylor residual difference and the recovered image obtained from the CIP implemented by the two methods are approximately the same on flux, with the $RMS$ of difference image as low as $10^{-5}$. The main difference is concentrated on the stronger point sources (as shown in Fig. \ref{fig:cipres}, difference), which is also on the dirty image (Fig. \ref{fig:wsimage}, mask of the difference).

\section{Discussion}
\label{sec:discussion}

\subsection{Reprojection}

\begin{figure*}
	\includegraphics[width=\textwidth]{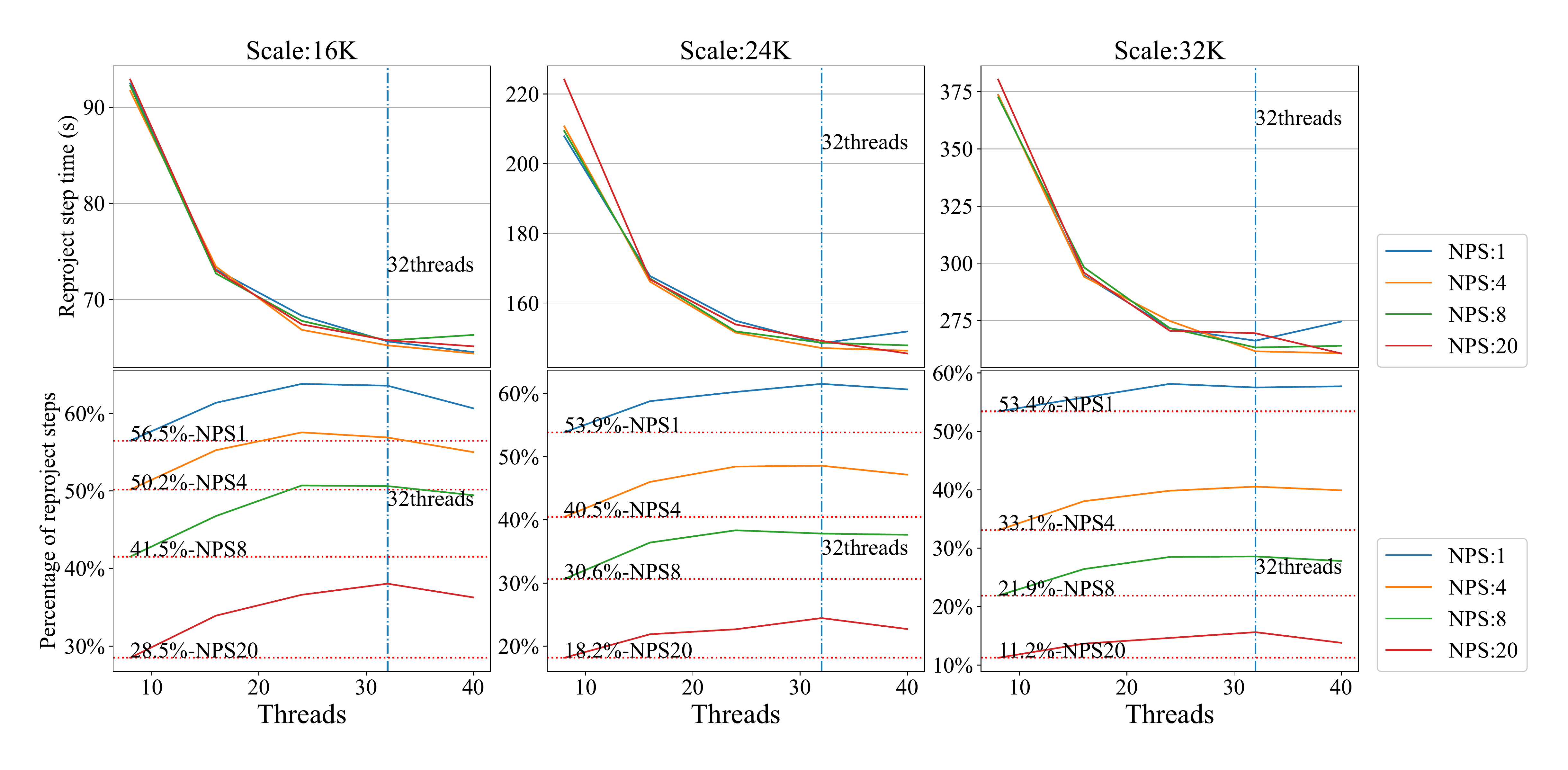}
    \caption{The figure shows the runtime and ratio of the reproject step of WS-Snapshot for single time slice with different parameters}
    \label{fig:per_reproject}
\end{figure*}
Reprojection plays a significant role in the overall WS-Snapshot implementation. Its performance also directly affects the performance of WS-Snapshot. However, the latest reproject package (version 0.8) cannot process a large-size image of 32K up. Meanwhile, the reproject package is a serial implementation that significantly decreases the WS-Snapshot performance. 

We optmised the ``efficient\_pixel\_to\_pixel\_with\_roundtrip'' function in the reproject package using multiple threads and keep the same number of threads as the gridding step for comparability of evaluation.
With fewer than 32 threads, the time percentage of reproject increases with threads (as shown in Fig. \ref{fig:per_reproject}).

Although multi-threading technology has been used, 
in calculating a single slice with NPS=1, reproject takes up more than 50 percent of the computation time. The optimisation efficiency in the assessment has space for improvement. Currently, only the bottleneck step ("efficient\_pixel\_to\_pixel\_with\_roundtrip") is parallelised using python multi-threading, and the interpolation and other steps remain based on the serial implementation.

\subsection{Precision}
WS-Snapshot can obtain a high degree of accuracy.
Figure~\ref{fig:per_error_wplane} shows that in the best case (NPS=1), the $RMS$ of difference image is as low as $10^{-6}$, with the main differences from the image distortion due to the projection. In the worst case (NPS=20), the error in the difference image reaches $10^{-4}$, a result that is unacceptable. This one shows that the value of NPS should be small when high accuracy results are needed.

It is necessary to note that it is also very difficult to further improve the accuracy of Ducc.wgridder.
Ducc.wgridder can dynamically select the kernel, kernel support, and oversampling factor to obtain maximum performance, and we can specify the error \citep{RN278} compared to the direct Fourier transform by changing the epsilon parameter of Ducc.wgridder. However, adjusting this parameter in the Ducc.wgridder step of WS-Snapshot has no effect on the final results of WS-Snapshot because the main errors are caused by the image reprojection.

\subsection{Network Data Transmission}
Network transmission is an essential factor in the performance of distributed computing. The data that must be transferred between the various computing nodes in the CIP imaging computation is mainly visibility data and images. Image transfer is the main load in the computation, and the size of an image can reach 8GB at a 32K scale.

The visibility data is distributed as it is split by frequency first, then further in time slices. Therefore, each computing node needs to read about 223MB of visibility data in the CIP at a 32K scale.

After the computation is completed, we use the reduced form to implement image and visibility merging, which reduces the single node memory usage and increases speed.

\section{Conclusions}

After conducting various performance assessments, we conclude that WS-Snapshot has the following advantages in large-scale imaging. 

1) Capable of achieving demanding widefield imaging processing with the high accuracy required for scientific studies. The combination of W-stacking and Snapshot can guarantee sufficient accuracy in large-scale imaging processing (error range of $10^{-4}$ to $10^{-7}$). The central part of the image can be selected based on the FoV when high-precision imaging is required to avoid the errors caused by reprojections.

2) When using the fitted optimal plane, the value of w-range becomes smaller (see Figure~\ref{fig:fitwplane}), which significantly improves the computational performance for IW-stacking (Ducc.wgridder). At a scale like 32K, WS-Snapshot nearly halves the computation time.

3) Since the IW-stacking algorithm is splittable in u,v,w directions, WS-Snapshot can be computed for distributions of observations at different times. This makes sense for data processing on the scale of SKA. 

Overall, WS-Snapshot inherits the advantages of the IW-Stacking and Snapshot methods. It can guarantee sufficient imaging accuracy and significantly reduce the runtime at large scale imaging.

The current study also shows that the bottlenecks encountered in the computation of WS-Snapshot deserve to be studied in more depth. These include how to perform the projection computation more quickly and with high accuracy and improve the performance of large scale image computation.

\section*{Acknowledgements}

This work is supported by the National SKA Program of China (2020SKA0110300), the Joint Research Fund in Astronomy (U1831204, U1931141) under cooperative agreement between the National Natural Science Foundation of China (NSFC) and the Chinese Academy of Sciences (CAS), the Funds for International Cooperation and Exchange of the National Natural Science Foundation of China (11961141001), the National Natural Science Foundation of China (No.11903009). The Innovation Research for the Postgraduates of Guangzhou University (2021GDJC-M13). The research of OS is supported by the South
African Research Chairs Initiative of the Department of Science and Technology and National Research Foundation. 

We do appreciate all colleagues of SKA-SDP ORCA and Hippo team for valuable and helpful comments and suggestions. 
This work is also supported by Astronomical Big Data Joint Research Center, co-founded by National Astronomical Observatories, Chinese Academy of Sciences and Alibaba Cloud.



\section{Data Availability}

We cloned RASCIL from its official repository and further implemented the WS-Snapshot. 
All the program code and part of the test data are stored in Gitlab repository, the RASCIL with WS-snapshot link is \url{https://gitlab.com/ska-sdp-china/rascil.git}, the link of the experimentally optimized reproject is \url{https://gitlab.com/ska-sdp-china/reproject.git}.



\bibliographystyle{mnras}
\bibliography{wssnapshot} 




\appendix


\bsp	
\label{lastpage}
\end{document}